\documentclass[prl,twocolumn,showpacs]{revtex4}
\usepackage{amsmath}
\usepackage[dvips]{graphicx}

\begin{document}

\title{Direct measurement of decoherence for entanglement between a photon
and stored atomic excitation}
\author{H. de Riedmatten, J. Laurat, C.W. Chou, E.W.
Schomburg, D. Felinto and H.J. Kimble}
\affiliation{Norman Bridge Laboratory of Physics 12-33, California Institute of
Technology, Pasadena, California 91125, USA}

\begin{abstract}
Violations of a Bell inequality are reported for an experiment where one of
two entangled qubits is stored in a collective atomic memory for a
user-defined time delay. The atomic qubit is found to preserve the violation
of a Bell inequality for storage times up to 21 $\mu $s, $700$ times
longer than the duration of the excitation pulse that creates the
entanglement. To address the question of the security of
entanglement-based cryptography implemented with this system, an investigation of
the Bell violation as a function of the cross-correlation between the
generated nonclassical fields is reported, with saturation of the
violation close to the maximum value allowed by quantum mechanics.
\end{abstract}

\maketitle

Entanglement between light and matter enables nonclassical correlation
between flying and stored quantum states, and as such is a critical resource
for quantum information science \cite{zollerepjd}. Among the capabilities
enabled by atom-light entanglement are the teleportation of quantum states
of light to a quantum memory \cite{Sherson06} and the heralded entanglement
between remote atomic systems (\cite{duan2004,ChouNature} and references therein). Generally, light-matter entanglement provides an essential enabling building
block for applications such as scalable quantum networks and quantum
repeaters over large distances \cite{repeater_01,DLCZ_01}.
Beyond the pioneering demonstrations of violations of Bell inequalities by
photons spontaneously emitted in atomic cascades \cite%
{clauser, aspect}, recent experiments have explicitly demonstrated
entanglement between the polarization states of single photons and the
internal spin states of single trapped atoms \cite%
{blinov04,volz06}. On the other hand, the seminal work of Duan, Lukin, Cirac
and Zoller (DLCZ) \cite{DLCZ_01,DLCZpra} spurred intense experimental and
theoretical efforts related to the entanglement of single photons and
\textit{collective excitations} in atomic ensembles. Advances on this front
include the generation \cite{Kuzmich03,Eisaman04}, storage \cite%
{Chaneliere05,Eisaman05,felinto05}, entanglement \cite{ChouNature,Matsu2},
and transfer from matter to light \cite%
{chou04,balic05,black05,laurat06,matsukevich06single} of single collective
atomic excitations, as well as probabilistic entanglement between internal
atomic Zeeman states and photon polarization \cite{Matsukevich05}. 

The relevance of atom-light entanglement for quantum network applications
arises from the fact that the material qubit can be stored and later
converted to a photonic qubit, while preserving its coherence. However, in
all experiments to date along this line \cite{blinov04,volz06,Matsukevich05}%
, no direct study was made of the decoherence process in the qubit storage.
For experiments with collective atomic excitations, the entanglement was
demonstrated for short storage times (e.g., $500$ ns in \cite{Matsu2}),
comparable to the duration of the excitation pulse. For quantum memory applications, it is however clearly important that the storage time is much longer than the time needed to address the memory. Longer coherence times
for atomic ensembles in the single-excitation regime have been inferred from
the decay of cross-correlation functions for the emitted light \cite%
{felinto05,Chaneliere05,Eisaman05,Matsukevich05,matsukevich06single}, but
without direct measurements of the lifetime for entanglement.

In this Letter, we report the first direct measurement of decoherence for one
stored component of a Bell state in an atomic memory. Polarization
entanglement is generated in a probabilistic way between a photon and a
collective atomic excitation. After
a variable storage time $\tau $, the atomic qubit is transferred into
a photon, and the polarization correlation with the
initial photon is measured as a function of $\tau$. The violation of a Bell
inequality is observed for storage times up to $\tau \simeq 21\,\mu $%
s, $700$ times longer than the duration of the initial
excitation pulse ($30$ ns). In addition, for small $\tau =400\, \text{ns}$, we
measure the Bell parameter $S$ as a function of the normalized
cross-correlation $g_{12}$ between the initial and retrieved photons, thereby
addressing operationally the relationship between the nonclassical character
of the generated fields and the security of a quantum channel implemented
with these resources \cite{ekert91}. Our observations are made possible by
two advances, namely a large improvement of the quality of the
photon pairs emitted by the atomic ensemble \cite{laurat06}, and the implementation of conditional logic for the
generation and read-out of the stored qubit.

\begin{figure}[th]
\centerline{\includegraphics[width=8cm,angle=0]{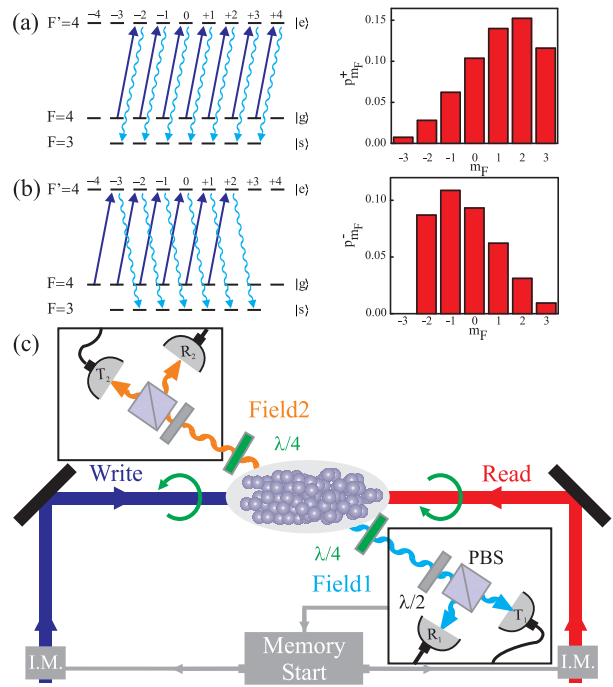}} \vspace{+0.15cm%
} 
\caption{(color online) (a) and (b). Relevant levels and decay paths from $|e\rangle $ to $|s\rangle $,
starting from an unpolarized Cs ensemble. The distributions $%
p_{m_{F}}^{+},p_{m_{F}}^{-}$ for populations in $|s\rangle $ from the two
possible decay paths are also shown. (c) Experimental setup. Write and read pulses are 
sent sequentially with 400 ns delay, until a detection in field $1$ occurs. This event triggers the \textquotedblleft Memory
Start\textquotedblright\ circuit, which stops
the write/read sequence for a programmable time $\protect\tau $ by way of
independent electro-optic Mach-Zehnder intensity modulators (I.M.). The read beam power is $\approx 150 \mu$W while the write beam power is much weaker ($\mu W$ range, see text). }
\vspace{-0.5cm}
\label{setup}
\end{figure}

Figure \ref{setup} provides an overview of our experiment, with (a, b)
illustrating\ the relevant pathways to generate entanglement probabilistically
between a photon and a collective atomic excitation, and (c) showing the
experimental setup. The optically-thick atomic ensemble is obtained from
cold Cs atoms in a magneto-optical trap (MOT). We call $\{|g\rangle
,|s\rangle ,|e\rangle \}$ the hyperfine levels $\{|6S_{1/2},F=4\rangle
,|6S_{1/2},F=3\rangle ,|6P_{3/2},F=4\rangle \}$,
respectively. With initially all atoms in
the ground state $|g\rangle $, a weak write pulse, detuned $10$ MHz
below the $g\rightarrow e$ transition and right circularly polarized ($%
\sigma ^{+}$), passes through the sample. With small probability $p$, an atom in
$|g,m_{F}\rangle $ undergoes spontaneous Raman scattering by way of the
excited state $|e,m_{F}^{\prime }=m_{F}+1\rangle $, and is thereby
transferred to $|s\rangle $ while emitting a photon (field $1$).
The spatial mode for field $1$ is defined by the backwards projection of our
imaging system into the ensemble \cite{laurat06}. The
transition $|g,m_{F}\rangle \rightarrow |s\rangle $ via $|e,m_{F}^{\prime
}=m_{F}+1\rangle $ proceeds by two different pathways: by emitting a $%
\sigma ^{+}$ polarized photon, arriving at $|s,m_{F}\rangle $, and
by emitting a $\sigma ^{-}$ photon, arriving then at $|s,m_{F}+2\rangle $%
, as in Figs.~1a and~1b, respectively. Also shown are the
expected distributions $p_{m_{F}}^{+},p_{m_{F}}^{-}$ of atoms in $|s\rangle $
as a result of $\sigma ^{+},\sigma ^{-}$ emission assuming an uniform initial distribution among the various $|g,m_{F}\rangle $. 
If the relevant emission processes are indistinguishable in all other degrees of
freedom, and field 1 is detected in a superposition state of $\sigma ^{+}$
and $\sigma ^{-}$, then the state of the excitation stored in $%
|s\rangle $ is projected into a coherent superposition of the mixed states shown in
Figs.~1a and~1b. In our experiment, the persistency of this projection is
evaluated as a function of the storage time $\tau$.

Before detection of the first photon, the joint state of the atom-light system for atoms initially in 
$|g,m_{F}\rangle $ can
be written as $\rho _{1a}=\left\vert
0\right\rangle \left\langle 0\right\vert +\left\vert \Phi _{1a}\right\rangle
\left\langle \Phi _{1a}\right\vert $, where the non-vacuum part is in the ideal case :
\begin{equation*}
\left\vert \Phi _{1a}\right\rangle =\sqrt{p}\Big[\cos \eta_{m_F} \left\vert 1
_{1}^{+},1 _{a}^{+}\right\rangle +\sin \eta_{m_F} \left\vert 1
_{1}^{-},1 _{a}^{-}\right\rangle \Big]+O(p).  \label{state}
\end{equation*}%
and $\left\vert 1_{1}^{\alpha}\right\rangle$ represents a photon in field 1 with a polarization $\sigma^{\alpha}$ and 
$\left\vert 1_{a}^{\alpha}\right\rangle$ the collective atomic states with one excitation as shown in Figs.~1a and~1b, for $\alpha =\{+,-\}$. The parameter $\eta_{m_F} $ is obtained from $\cos ^{2}\eta_{m_F} =
p_{m_{F}}^{+}/(p_{m_{F}}^{+}+p_{(m_{F}+2)}^{-})$. For the more general case where the initial state is an incoherent distribution of the various $|g,m_{F}\rangle $, the collective atomic states are mixed states and the global $\eta $ is obtained from $\cos ^{2}\eta =\sum
p_{m_{F}}^{+}/\sum (p_{m_{F}}^{+}+p_{m_{F}}^{-})$~\cite{Matsukevich05},
where for the case of Cs atoms, $\eta =0.86\times \pi /4$. Note that the vacuum part in $\rho_{1a}$ also contains all the light emitted by the ensemble which is not collected in the single mode of our imaging system \cite{DLCZpra}.
By sending a strong read pulse, $\sigma ^{-}$ polarized
with respect to the atoms and resonant with the $%
s\rightarrow e$ transition, the atomic qubit can be transferred efficiently
into a single photon (field 2). Field $2$ is emitted into a
well-defined spatial mode \cite{laurat06} and with polarization orthogonal
to field $1$ thanks to a collective enhancement effect \cite%
{DLCZ_01,felinto05}. Hence, the atomic qubit is mapped onto a photonic qubit with, for each $m_F$, the atomic state 
$\left\vert 1_{a}^{\alpha}\right\rangle$ being mapped onto a photonic state $\left\vert 1_{2}^{-\alpha}\right\rangle$ in field 2.

Returning to the experimental setup in Fig. \ref{setup}c, we carry out the
excitation and retrieval in a cyclic fashion. At a frequency of $40$ Hz, the
MOT magnetic field is switched off for $6$ ms. After waiting for $3.8$ ms
for the magnetic field to decay \cite{felinto05}, a sequence of $1100$
trials of duration $2$ $\mu $s begins. For each trial, the atoms are
initially prepared in $|g\rangle $ with $1$ $\mu $s of
repumping light. Write and read pulses, each of 30 ns
duration, are mode matched and counter propagate through the ensemble
with a beam waist $\simeq 200$ $\mu $m. Fields $1$ and $2$ are collected with
a $3^{\circ }$ angle relative to the write and
read beams \cite{balic05,Matsukevich05,laurat06}, and with a waist in the
MOT $\simeq 50$ $\mu $m. They are then directed to $\lambda/4$ 
plates that map circular to linear polarization, and then to
rotatable polarizers with angles $\theta _{1}$, $\theta _{2}$,
each consisting of a $\lambda/2$ plate and a polarization beam splitter (PBS). The two outputs of the PBSs are then
coupled to single-mode optical fibers and sent to silicon avalanche
photodiodes, denoted by ($T_{1,2},R_{1,2}$) for the transmitted
and reflected outputs, for fields $1$,$2$, respectively. Before
detection, field $1$ is sent through a paraffin-coated vapor cell containing Cs atoms in the $|g\rangle $ state, used as a frequency filter to reduce uncorrelated background
events \cite{laurat06}. Finally, the detection electronic signals are sent to a data acquisition card, where they are
time-stamped and recorded for later analysis.

Before studying the storage process, a first characterization
 at short storage times $\tau =400$ ns is obtained by way of
the correlation function $E(\theta _{1},\theta _{2})$ defined by
\begin{eqnarray}
E(\theta _{1},\theta _{2})=\frac{%
C_{T_{1}T_{2}}+C_{R_{1}R_{2}}-C_{T_{1}R_{2}}-C_{R_{1}T_{2}}}{%
C_{T_{1}T_{2}}+C_{R_{1}R_{2}}+C_{T_{1}R_{2}}+C_{R_{1}T_{2}}}\text{ .}
\label{corr}
\end{eqnarray}%
Here $C_{T_{1}T_{2}}$ gives the number of coincidences
between detectors $T_{1}$ and $T_{2}$ for the angles $\theta _{1}$ and $%
\theta _{2}$. For the generation of photon pairs from an
atomic ensemble, the \textquotedblleft quality\textquotedblright\ of the pairs depends on the intensity of the excitation (writing) laser \cite%
{laurat06}, as in parametric downconversion. For low
excitation intensity, the non-vacuum part is well
approximated by a photon pair, but as the excitation increases, the higher
order terms can no longer be neglected. We assess the contributions of these
higher order terms by way of measurements of the normalized cross
correlation function $g_{12}$ between the two fields, where $%
g_{12}=p_{12}/(p_{1}p_{2})$, with $p_{12}$ as the joint probability for
detection events from fields $1,2$ in a given trial, and $p_{i}$ as the
probability for unconditional detections in field $i$.
For our system  $g_{12}>2$ is a strong indication of a nonclassical state of
light for the two fields \cite{Kuzmich03,felinto05}.

\begin{figure}[th]
\vspace{-0.2cm}
\centerline{\includegraphics[width
=7.8cm,angle=0]{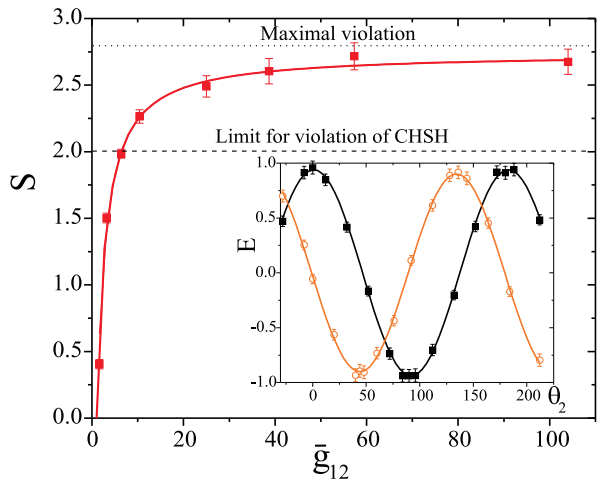}} \vspace{-0.3cm} 
\caption{(color online). Measurement of the Bell parameter S as a function
of the average value for the normalized cross correlation function $\bar{g}%
_{12}$. The fitted $S_{max}$ is $2.74\pm 0.04$. 
Inset: Measurement of the correlation function $E(\protect\theta %
_{1},\protect\theta _{2})$  as a function of
 $\protect\theta _{2}$, for an average $%
\bar{g}_{12}=57$. Filled squares shows the fringe for $\protect\theta _{1}$ =$%
0^{\circ }$ ($V=0.94\pm 0.02$), while open circles correspond to $\protect\theta _{1}=-45^{\circ }$
($V=0.90\pm 0.02$).}
\vspace{-0.3cm}
\label{Bellg12}
\end{figure}

An example of $E(\theta _{1},\theta _{2})$ is shown
in the inset of Fig. \ref{Bellg12}, where $E$ is
displayed as a function of $\theta _{2}$ for two different values of $\theta
_{1}=0^{\circ},-45^{\circ}$ corresponding to the projection of photon $1$ on 
bases separated by $45^{\circ}$. These curves were
taken for an average value of $\bar{g}_{12}=57$, measured at the point of
maximal correlation ($\theta _{1}=\theta _{2}=0^{\circ}$), so that the
transmitted field $1$ (field $2$) is $\sigma ^{-}$ ($\sigma ^{+})$
polarized. The value $\bar{g}_{12}$ is the average between the two
polarization processes described in Figs. 1a and 1b, i.e., where fields $1$ and $2$ are detected with $\sigma ^{+}$ and $\sigma ^{-}$
polarization ($\sigma _{1}^{+},\sigma _{2}^{-}$), and vice versa ($\sigma
_{1}^{-},\sigma _{2}^{+}$).
From measurements of $E(\theta _{1},\theta _{2})$, it is possible to determine the Bell parameter $S=E(\theta _{1},\theta _{2})+E(\theta _{1}^{\prime },\theta _{2})+E(\theta
_{1},\theta _{2}^{\prime })-E(\theta _{1}^{\prime },\theta _{2}^{\prime })%
$ \cite
{clauser}. 
We use the canonical settings $\theta _{1}=-22.5^{\circ},\theta _{1}^{\prime
}=22.5^{\circ},\theta _{2}=0^{\circ},\theta _{2}^{\prime }=45^{\circ}$,
which give $S=2\sqrt{2}$ for an ideal entangled state of two qubits. This corresponds to the maximal violation of the Bell-CHSH
inequality $\left\vert S\right\vert \leq 2$ \cite{clauser}.

Our results for $S$ for fixed $\tau =400$ ns are displayed in
Fig. \ref{Bellg12} as a function of $\bar{g}%
_{12}$ as the write beam power is varied from $\approx$ 0.3 to 20 $\mu W$. Violations of the CHSH
inequality are evident for large $\bar{g}%
_{12}$, but are lost as $\bar{g}_{12}$ is reduced. This loss is due to the
higher order terms, which act as
background noise that tends to reduce the visibility $V$ of the 
fringes in $E(\theta _{1},\theta _{2})$. Since $p_{1}p_{2}$ gives a good estimation for the uncorrelated
background, $V$ can be approximated by:
\begin{equation}
V\simeq \frac{p_{12}-p_{1}p_{2}}{p_{12}+p_{1}p_{2}}=\frac{g_{12}-1}{g_{12}+1}%
\text{ .}  \label{V}
\end{equation}%
The solid line in Fig. \ref{Bellg12} is a fit with the expression $%
S=S_{max}V $, where $S_{max}$ is the maximal
possible violation \cite{marcikic04}, and $V$ is given by Eq.(\ref{V}). Consistent with Eq.(\ref{V}), $%
S$ reaches a plateau for high $\bar{g}_{12} $ with the fitted value $%
S_{max}=2.74\pm 0.04$ close to the maximal violation $2.79$ expected for a
process with $\eta =0.86\times \pi /4$ \cite%
{Matsukevich05}. Our maximum measured value is $S=2.7\pm 0.1$, near the
maximal violation and representing a violation by $7$ standard deviations of
the CHSH inequality. Also of note is that the threshold $\left\vert
S\right\vert =2$ for violation of the CHSH inequality occurs for $\bar{g}%
_{12}\simeq 7$. Although there has been tremendous progress in the
achievable value of $g_{12}$ in recent years \cite{laurat06,matsukevich06single}, 
no study has previously investigated the relationship of
the quantum correlations represented by $g_{12}$ with the requirements for
quantum network applications (e.g., violation of a Bell-inequality for the
security of entanglement-based quantum cryptography \cite%
{ekert91,norbert}). The measurements in Fig. \ref{Bellg12} represent the first step
to quantify this connection.

We next investigate the time interval $\tau $ over which excitation can be
stored in the atomic memory while still preserving sufficient coherence for
violation of the CHSH inequality. For
this study, the period of the trials must be increased to beyond the
decoherence time for the stored qubit (up to $\tau _{\max }=40$ $\mu s$
in our case). Because the success probability $p_{1}$ for a detection event
from field $1$ is necessarily small ($p_{1}\simeq 10^{-4}$ for $%
\bar{g}_{12}\simeq 60$), the time $t_{s}$ required for successful detection
becomes long ($t_{s}\simeq \tau _{\max }/p_{1}$), leading to
prohibitively low count rates if the experiment were to be conducted in the
usual cyclic fashion. To circumvent this problem, we have developed a
control system that stops all light pulses for a programmable time $\tau $
conditioned upon a detection event for field $1$, before the read pulse is fired. Operationally, the repetition rate
for our experiment is thereby increased by more than a factor of $20$ as
compared to usual (unconditional) cycling.

\begin{figure}[th]
\vspace{+2.2cm} \centerline{%
\includegraphics[width=9.5cm,angle=0]{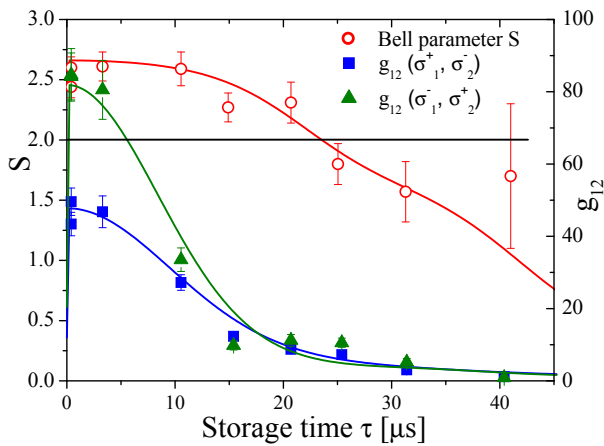}} \vspace{-3.2cm}
\caption{(color online) Measurement of the Bell parameter S (open circles)
and of $g_{12}$ for the two different polarization configurations (filled
symbols) as a function of $\tau$. To increase
the repetition rate, repumping light has not been used between trials and
the period has been shortened to 1.45 $\protect\mu $s, such that 1400 trials
can be performed in a measurement period.}
\vspace{-0.6cm}
\label{Bellsto}
\end{figure}

The open circles in Fig. \ref{Bellsto} give the results for our measurement
of $S$ as a function of $\tau $. With the same set of angles as in Fig. \ref%
{Bellg12}, we find $S=(2.31\pm 0.17)>2$ at $\tau =20.7 $ $\mu $s delay.
Hence, the storage of the atomic qubit preserves the violation of the CHSH
inequality up to $\tau \simeq 21$ $\mu s$, corresponding to $%
4$ km propagation delay in an optical fiber. The principal cause for the
decay of $S$ with increasing $\tau $ is the residual magnetic field that
inhomogeneously broadens the ground state levels $|g,m_{F}\rangle $, $%
|s,m_{F}^{\prime }\rangle $, as was studied in detail in Ref. \cite%
{felinto05}.

To substantiate this claim, we display in Fig. \ref{Bellsto} measurements of
$g_{12}$ for the two different polarization configurations $(\sigma
_{1}^{+},\sigma _{2}^{-})$ and $(\sigma _{1}^{-},\sigma _{2}^{+})$ taken at
the same time as those for $S$. $g_{12}$ likewise exhibits decay with
increasing $\tau $ that we investigate by applying the model
introduced in Ref. \cite{felinto05}. We calculate the joint probability $p_{12}^{th}(\tau)$ to generate a pair of photons
in fields $1,2$. We then compare the quantity $p_{12}(\tau )=\xi
p_{12}^{th}(\tau )$ to the measured $g_{12}(\tau )$ by way of a single
overall scaling parameter $\xi $ for all $\tau $, for each polarization
configuration, resulting in the solid lines in Fig. \ref{Bellsto}. The
observed decay is consistent with an inhomogeneity of the Zeeman splitting
across the ensemble described by the parameter $K=\mu _{B}g_{Fg}Lb/h$
where $L$ is the ensemble length, $b$ the residual magnetic field
gradient, and $g_{Fg}$ the Land\'{e} factor. The fits in Fig. \ref{Bellsto}
are for $K=12$ kHz for the two polarization configurations, which is
consistent with the linewidth of the ground state determined independently
by stimulated Raman spectroscopy \cite{felinto05}. We have no definitive
explanation for the different measured values of $g_{12}$ at $\tau =0$ for
the two configurations but suggest that this difference might
be due to different backgrounds (consistent with the different $\xi $
values for the two curves). The conditional probability $p_c$ to
detect a photon in field $2$ conditioned on a detection in field $1$ follows the decay of $g_{12}$, starting at $6\%$ for $\tau =0$ and falling
to $0.7\%$ for $\tau =20.7$ $\mu $s.

From the theoretical curves for $g_{12}$ in Fig. \ref{Bellsto}, we obtain a
prediction for the decay of the Bell parameter $S$ also shown by a solid curve
in Fig. \ref{Bellsto}. Explicitly, we assume as before that $S=S_{max}V$
\cite{marcikic04}, with the visibility $V$ calculated from the average $\bar{%
g}_{12}$ of the modeled decay for the two polarization configurations
by way of Eq. \ref{V} and the $S_{max}=2.74$ obtained from the fit in Fig. \ref%
{Bellg12}. The agreement between this simple model and our measured values
of $S$ indicates that the principal cause of decoherence for the Bell
inequality violation is well understood. 

In summary, we have described a Bell experiment based on probabilistic
entanglement between a photon and a collective atomic excitation, where one
of the qubits is stored in an atomic ensemble before being transfered to a
single photon. Within the setting of the realization of scalable quantum networks via the protocol of DLCZ \cite{DLCZ_01}, we have
presented the first measurements to explore the connection between
traditional field correlations as expressed by $g_{12}$ and entanglement as
represented by the Bell parameter $S$. The storage of the matter qubit leads
to a violation of the CHSH inequality for storage times up to $21$ $\mu $s,
with the mechanism for decoherence identified theoretically. For scalable quantum networks with a number of nodes $>>$ 2, the coherence time should be much larger than the time needed to create the entanglement with high probability, which remains an experimental challenge \cite{felinto05}. Beyond the
setting of the DLCZ protocol, our results represent the first direct
measurements of the decoherence in the storage of a matter qubit in an
atom-light entanglement experiment.

We gratefully acknowledge our ongoing collaboration with Dr. S. J. van Enk.
This work was supported by the DTO of the
DNI and by the NSF.
J.L. acknowledges financial support from the EU (Marie Curie
Fellowship), and D.F. from CNPq (Brazilian agency).

\end{document}